\documentclass[12pt]{article}
\usepackage[body={17.5cm, 22cm},right=2cm]{geometry}             

\newenvironment{changemargin}[2]{%
  \begin{list}{}{%
    \setlength{\topsep}{0pt}%
    \setlength{\leftmargin}{#1}%
    \setlength{\rightmargin}{#2}%
    \setlength{\listparindent}{\parindent}%
    \setlength{\itemindent}{\parindent}%
    \setlength{\parsep}{\parskip}%
  }%
  \item[]}{\end{list}}
\usepackage{axodraw4j}
\usepackage{pstricks}
\usepackage{color}
\usepackage{epsfig}

\usepackage{graphicx}
\usepackage{setspace}
\usepackage{amssymb}
\usepackage{listings}
\usepackage{epstopdf}
\usepackage{float}
\usepackage{cite}
\usepackage{fancyheadings}
\usepackage{color}
\usepackage{amsmath}
\usepackage{ dsfont }
\usepackage{hep}
\usepackage{subfigure}

\begin{document}
\begin{center}
\begin{changemargin}{-0.5cm}{-0.5cm}
{\Large\bf \quad Graviton Phenomenology of Linear Dilaton Geometries}\\
\end{changemargin}
\vspace{0.5cm}
{ \large
Masha Baryakhtar\footnote{email: mbaryakh@stanford.edu}\\ 
\vspace{.45cm}
{\small  \textit{Stanford Institute for Theoretical Physics, Department of Physics, \\Stanford University, Stanford, CA 94305 }}\\ }
\end{center}
\begin{center}
\vspace{.5cm}
\begin{abstract}
  Five-dimensional geometries with a linearly varying dilaton
  background arise as gravity duals of TeV Little String Theories
  (LSTs) and provide a solution of the hierarchy problem through extra
  dimensions. The unique Kaluza-Klein graviton spectrum has a mass gap
  on the order of the dilaton slope followed by a closely spaced
  discretum of states. We study in detail the graviton phenomenology
  in this scenario, allowing the dilaton slope to vary from the MeV to
  the TeV scale. When the dilaton slope is large enough so that
  individual KK resonances can be resolved at the LHC, several of them
  can be discovered simultaneously and allow for the linear dilaton
  geometry to be uniquely identified. For much smaller values of the
  dilaton slope, the LHC signatures become similar to the 5-d ADD
  scenario while relaxing the astrophysical and experimental
  constraints. Due to the mass gap, the KK modes are produced on-shell
  and decay inside the LHC detector, modifying the diphoton and
  dilepton spectra at large invariant mass. Finally, we perform a
  similar analysis for the low curvature RS geometry. We present
  experimental limits and calculate the ultimate reach of a 14 TeV LHC
  for all the above scenarios.

\end{abstract}
\end{center}

\newpage
\tableofcontents

\newcommand{\fig}[1]{figure~\ref{#1}}
\newcommand{\hc}{{\rm h.c.}}
\newcommand{\ds}{\displaystyle}
\newcommand{\eqr}[1]{(\ref{#1})}
\newcommand{\tr}{\mathop{\rm tr}}
\newcommand{\uone}{{U(1)}}
\newcommand{\su}[1]{{SU(#1)}}
\newcommand{\stu}{\su2\times\uone}
\newcommand{\be}{\begin{equation}}
\newcommand{\ee}{\end{equation}}
\newcommand{\bal}{\begin{align}}
\newcommand{\eal}{\end{align}}
\newcommand{\bp}{\begin{picture}}
\newcommand{\ep}{\end{picture}}
\renewcommand{\thefootnote}{\alph{footnote}}
\def\spur#1{\mathord{\not\mathrel{#1}}}
\def\lta{\mathrel{\displaystyle\mathop{\kern 0pt <}_{\raise .3ex
\hbox{$\sim$}}}}
\def\gta{\mathrel{\displaystyle\mathop{\kern 0pt >}_{\raise .3ex
\hbox{$\sim$}}}}
\newcommand{\sechead}[1]{\medskip{\bf #1}\par\bigskip}
\newcommand{\ba}[1]{\begin{array}{#1}\ds }
\newcommand{\cra}{\\ \ds}
\newcommand{\ea}{\end{array}}
\newcommand{\forto}[3]{\;{\rm for}\; #1 = #2 \;{\rm to}\; #3}
\newcommand{\for}{\;{\rm for}\;}
\newcommand{\cross }{\hbox{$\times$}}
\newcommand{\ol}{\overline}
\newcommand{\g}[1]{\gamma_{#1}}
\newcommand{\half}{{1\over 2}}
\newcommand{\del}{\partial}
\newcommand{\grad}{\vec\del}
\newcommand{\real}{{\rm Re\,}}
\newcommand{\imag}{{\rm Im\,}}
\newcommand{\mev}{\,\mathrm{MeV}}
\newcommand{\gev}{\,\mathrm{GeV}}
\newcommand{\tev}{\,\mathrm{TeV}}
\newcommand{\ev}{\,\mathrm{eV}}
\newcommand{\kev}{\,\mathrm{KeV}}

\newcommand{\ifb}{\,\mathrm{fb}^{-1}}
\newcommand{\ipb}{\,\mathrm{pb}^{-1}}

\def\cO{\mathcalL{O}}
\def\cL{\mathcal{L}}
\def\half{\frac{1}{2}}
\def\l{\left}
\def\r{\right}
\def\p{\partial}
                                      
\newcommand{\s}[1]{\slash{#1}}
\section{Introduction}

One of the most compelling connections between the weak scale and the
Planck scale appears in the framework of extra dimensions, in which the
fine-tuning problem becomes a question of dynamics and geometry
\cite{ArkaniHamed1998, Randall1999a}. In particular, in string theory the
$4$-dimensional Planck scale is not uniquely set by the string scale $M_s$,
but also depends on the string coupling $g_s$ and the internal volume of the six
compactified dimensions $V_6$,
\begin{align}
  \label{eq:36}
  M_{Pl}^2 = \frac{1}{g_s^2}M_s^8V_6.
\end{align}
Thus, there are two avenues to explain the weakness of gravity while
keeping the string scale near a TeV: the large
size of extra dimensions $V_6$, which dilute the strength of gravity
on the infrared brane \cite{ArkaniHamed1998}, or the smallness of the string coupling
constant $g_s$. The latter can be achieved in a controlled limit in
the case of Little String Theories (LSTs) \cite{Antoniadis2001}. 

Little String Theories are 6-d strongly coupled non-Lagrangian
theories generated by stacks of NS5 branes. The LSTs are dual to a
local 7-d theory with a linearly varying dilaton background in the
infinite 7th dimension. To obtain a finite 4-d Planck mass we use a
5-d compactification of this geometry introduced in
\cite{Antoniadis2011}, in which the infinite linear dilaton extra
dimension is terminated with two branes to a size $r_c$ and two more
extra dimensions are compactified on a smaller scale. Existing UV
motivated extra-dimensional frameworks are the ADD flat compactification
and the near-horizon limit of stacks of D3 branes which gives
rise to Randall-Sundrum (RS) geometry \cite{ArkaniHamed1998,
  Randall1999a}. The Linear Dilaton background, as the near horizon
limit of stacks of NS5 branes, presents a third such possibility.

The dilaton gravity theory can be approximated by the bulk action
\begin{align}
  \label{eq:44}
  S = \int d^5 x\sqrt{-g}\,e^{-\frac{\Phi}{M_5^{3/2}}}\l(M_5^3 R +
  (\nabla \Phi)^2 -\Lambda \r),
\end{align}
with the Standard Model on the $x_5\equiv z=0$ brane and the Planck
brane at $z=r_c$. The linear dilaton solution is achieved by imposing
the linearly varying background $\Phi/M_5^{3/2} = \alpha |z|$, with
the bulk metric solution 
\begin{align}
  \label{eq:45}
  ds^2 = e^{-\frac{2}{3}\alpha|z|}(\eta_{\mu\nu}dx^{\mu}dx^{\nu}+dz^2).
\end{align}

This geometry gives rise to a gapped Kaluza-Klein (KK) graviton spectrum with a massless
graviton and the first mode of the KK tower appearing at ${|\alpha|}/{2}$:
\begin{align}
  \label{eq:38}
  m_n&=\frac{|\alpha|}{2}\sqrt{1+\frac{4n^2\pi^2}{(\alpha r_c)^2}}.
\end{align}
To further study the observable consequences of the gravitational
sector we adopt the 5-d linear dilaton geometry without considering
other states such as strongly coupled little string excitations that
may be present in the LST.

In the Little String Theory context, the linear dilaton slope $\alpha$
is set by the number $ N$ of NS5 branes and for small $N$ is naturally
close to the 5-d Planck scale $M_5$. In this study we consider the
slope $\alpha$ as a free parameter which controls the mass scale of
the KK modes. We compute the detailed experimental consequences at high
curvature, $|\alpha|\sim M_5$; this scenario has the distinctive
signature of many closely spaced resonances with decay modes to jets,
leptons, and photons.

In addition, the limit of a low curvature 5th dimension reopens the
exciting possibility of discovering one large `flat' extra dimension
at the Large Hadron Collider (LHC). High-energy collisions at the LHC
are insensitive to the small curvature, so the phenomenology is very
similar to that of ADD with one extra dimension. With zero curvature
one extra dimension runs into severe astrophysical limits in addition
to being ruled out by everyday observation: the size would need to be close to the
Earth-Sun distance in size to result in a TeV-scale 5-d Planck mass
\cite{ArkaniHamed1998}. A low curvature possibility was previously
considered in \cite{Giudice2005}: an RS-type extra dimension with
small $60\mev$ curvature appears as flat at the
LHC but reduces the physical size of the extra dimension needed to
solve the hierarchy problem and avoids astrophysical bounds that rule
out one large extra dimension. The linear dilaton geometry provides an
alternative framework for this scenario.

This paper is organized as follows. In Section \ref{Section2} we
analyze the Kaluza-Klein mass spectra and the emergence of the
hierarchy as a function of geometry in Randall-Sundrum and Linear
Dilaton (LD) type backgrounds. Next, we examine the theory in several
different regions of parameter space. We discuss the case of high
curvature and corresponding LHC bounds and reach in Section~\ref{LargeAlpha}. In
Section \ref{LowAlpha}, we focus on the ``large $N$'' limit, treating
the curvature $\alpha$ as a free parameter. We present relevant
astrophysical and laboratory constraints, deriving limits on $\alpha$
and the 5-dimensional Planck scale $M_5$. In Section \ref{lhc} we compute
the LHC signals and study the current data for the low curvature
$|\alpha |\ll M_5$ limit. We obtain experimental bounds and comment on
optimal LHC searches and future reach of the collider.  In Section
\ref{LowRS} we discuss the low curvature RS geometry and apply
experimental constraints to this spectrum. In Section
\ref{conclusions} we summarize our findings and present searches to
distinguish the Linear Dilaton geometry with experiment, as well as
comment on its limitations in the context of Little String Theory at a
TeV.

\section{Hierarchy from Geometry}\label{Section2}

In both Randall-Sundrum and Linear Dilaton geometries, the curvature in
the 5th dimension is crucial to both the low-energy phenomenology of
the theory and to understanding the solution to the hierarchy
problem. In this section we relate the relevant quantities in the two
theories to develop intuition for curved geometries and their
phenomenology.

The well-known RS metric is given by 
 \begin{align}
   \label{eq:35}
   ds_{RS}^2 = e^{-2k|y|}\eta_{\mu\nu} dx^{\mu}dx^{\nu}+dy^2,
 \end{align}
 where $x^{\mu}$ are the 4-dimensional coordinates and $y$ is the physical
 distance along the extra dimension. The finite extra dimension is
 bounded by a UV brane at $y=0$ and IR brane at $y=b$, and all scales
 are referenced to the UV brane. 

 The conformally flat form of \eqref{eq:35} follows from the coordinate
 transformation $\dfrac{dz}{1+k|z|}~=~dy$,
\begin{align}
  \label{eq:3}
    ds_{RS}^2 = \frac{1}{(1+k|z|)^2}(\eta_{\mu\nu} dx^{\mu}dx^{\nu}+dz^2).
\end{align} 
We can compare this conformally flat form directly to the linear
dilaton metric \cite{Antoniadis2011},
\begin{align}
  \label{eq:39}
  ds_{LD}^2 = e^{-\frac{2}{3}\alpha |z| }(\eta_{\mu\nu} dx^{\mu}dx^{\nu} + dz^2),
\end{align}
where as in RS the extra dimension $z$ is finite and compactified on
an orbifold with $Z_2$ symmetry. In this case all parameters are
defined with respect to the IR brane at $z=0$ which contains the
Standard Model fields; the UV brane is at $z=r_c$. For a given 5-d
Planck mass $M_5$ and slope $\alpha$ of the dilaton field the size of
the extra dimension $r_c$ is fixed by the hierarchy between the 4-d
$M_{Pl}$ and 5-d Planck mass scales:
\begin{align}
  \label{eq:7}
  M_{Pl}^2 = 2\int_0^{r_c}\,dz\,e^{-\alpha|z|}M_5^3 =
  -2\frac{M_5^3}{\alpha}(e^{-\alpha r_c} -1).
\end{align}
This relation also fixes $\alpha < 0$. For typical values $M_5=1\tev$
and $|\alpha| = 1\gev$, we see that $|\alpha r_c|=60$ and the quantity
$|\alpha r_c|$ depends only logarithmically on $\alpha$ and $M_5$ for
a fixed value of the hierarchy between $M_{Pl}$ and $M_5$.

In the RS geometry \eqref{eq:3}, the change of mass scales along the
bulk can be understood as the running of the cutoff in the dual CFT
theory \cite{Arkani-Hamed2001}. The model with a linearly varying
dilaton background is not scale invariant and the dual theory is
strongly coupled, so a similar analysis is not possible
here. Nevertheless, important analogies to RS phenomenology remain.

The warped geometry provides an explanation for the hierarchy between
the weak scale and the Planck mass. We can see that mass scales are
warped down on the IR brane relative to the UV brane in both RS and LD
by considering the behavior of the Higgs vacuum expectation value
(VEV). In RS, the induced 4-d metrics at the IR and UV branes are
\begin{align}
  \label{eq:50}
 \mathrm{IR}:\, \l. g_{\mu\nu}\r|_{y=b}^{RS} = e^{-2kb}\eta_{\mu\nu}\qquad  \mathrm{UV}:\,  \l. g_{\mu\nu}\r|_{y=0}^{RS} = \eta_{\mu\nu}.
\end{align}
For a comparison to the LD geometry we rewrite the metric \eqref{eq:39} in
physical coordinates, $dy= e^{-\frac{1}{3}\alpha z} dz$,
\begin{align}
  \label{eq:40}
  ds_{LD}^2 =\l(1+\frac{|\alpha y|}{3}\r)^2\eta_{\mu\nu} dx^{\mu}dx^{\nu} + dy^2.
\end{align}
We see that the induced metrics on the two branes in LD are
\begin{align}
  \label{eq:49}
  \mathrm{IR}:\, \l. g_{\mu\nu}\r|_{y=0}^{LD} =\eta_{\mu\nu} \qquad  \mathrm{UV}:\,  \l. g_{\mu\nu}\r|_{y=y_0}^{LD} =\l(1+\frac{|\alpha
  y_0|}{3}\r)^2\eta_{\mu\nu}.
\end{align}
Since the induced metric differs between branes in both cases, to
compare the behavior of the mass scales we redefine the Higgs field to
be canonically normalized, leading to a warped value of the VEV
\cite{Randall1999a}. In RS the warping is exponential, $v_{IR} = e^{-kb}\,\,v_{UV}$, and in linear
dilaton it is a power law, $v_{IR} = \l(1+{|\alpha y_0|}/{3}\r)^{-1}\,v_{UV}$.  In
general, the Higgs VEV is warped down on the IR brane relative to the
UV so the weak mass scale is suppressed compared to $M_{Pl}$.

Along with the hierarchy, geometry also determines the 4-d KK
graviton mass spectra. This is clear if we rescale the metric
perturbation $h_{\mu\nu}$ to write the graviton equation of motion in
the form of a Schrodinger equation. In the LD model, the curvature
acts as a mass term: the spectrum corresponds to that of a particle
with mass $|\alpha|/2$ in a box of size $r_c$ \cite{Antoniadis2011},
\begin{align}
  \label{eq:21}
\eta^{MN}\partial_{M}\partial_{N}\tilde{h}_{\mu\nu}^{(n)}-\frac{\alpha^2}{4}\tilde{h}_{\mu\nu}^{(n)}
  = 0.
\end{align}
In RS on the other hand, the masses of the KK modes correspond to a
massless particle in a potential: they are determined by Bessel
function zeros corresponding to solutions of the potential $V(y)$ \cite{Randall1999a},
\begin{align}
  \label{eq:37}
\eta^{MN}\partial_{M}\partial_{N}\tilde{h}_{\mu\nu}^{(n)}-V(y)\tilde{h}_{\mu\nu}^{(n)}
  = 0.
\end{align}
The physical quantity that determines the graviton spectrum is the
time it takes for light to travel from the UV to the IR brane. In the
case of RS the time is at the TeV scale,
\begin{align}
  \label{eq:41}
  \Delta t^{\,null}_{RS}  = \frac{e^{kb}}{k}
\end{align}
For LD, the time it
takes for light to travel from UV to IR brane is just
\begin{align}
  \label{eq:41}
  \Delta t^{\,null}_{LD}  = r_c
\end{align}
For $|\alpha| = 1\tev$, $r_c=10^{-17}$ m, and even for a very low
curvature $|\alpha| =20\mev$, the proper size is small, $r_c=10^{-12}$
m. Compare this to the physical size of the 5th dimension: in physical
coordinates \eqref{eq:40} the proper distance $y_0$ between the branes
is of order $3|\alpha|^{-1}e^{\frac{|\alpha r_c|}{3}}$, which can be as
large as $10\, \mu$m for $|\alpha| = 20\mev$. However, the mass
spectrum \eqref{eq:38} is determined by the inverse time it takes
light to travel between the branes---in this case $r_c^{-1}$---as well
as the curvature $\alpha$, and not the proper length of the extra
dimension $y_0$. Since the graviton KK mass sets the scale of
modification of the gravitational $1/r^2$ law, laboratory and
astrophysical bounds that rule out a flat fifth dimension
\cite{Hall1999,Cullen1999,Hanhart2001,Hannestad2003,Adelberger2007}
have limited application to this scenario.

\section{High Curvature ($|\alpha|\sim M_5$) }\label{LargeAlpha}
\subsection{Phenomenology}

In the 5-d linear dilaton compactified geometry, Standard Model fields
are localized on the IR brane and the graviton propagates in the
bulk. The equation of motion in the bulk corresponds to that of a free
massive particle, which when compactified gives rise to a gapped mass
spectrum in 4-d. The details of the compactification can be found in
\cite{Antoniadis2011}.  The mass spectrum is unique: the mass
gap is set by the curvature $\alpha$, with the first massive state at
$m_1=|\alpha|/2$, followed by a discretum of states spaced by
$r_c^{-1}$ with masses proportional to the curvature $\alpha$,
\begin{align}
  \label{eq:30}
  m_n=\frac{|\alpha|}{2}\sqrt{1+\frac{4n^2\pi^2}{(\alpha r_c)^2}}.
\end{align}
Unlike condensed matter systems with ubiquitous mass gaps, the
spectrum is very rare in gravity since gravitational interactions
cannot be screened in positive-energy systems. In this case, a linearly
varying dilaton provides a quadratic potential term for the modes in
the extra dimension, resulting in a KK masses resembling the energy
spectrum of a `massive particle in a box' in 4-d.

Each KK mode couples to all Standard Model fields equally through the
stress-energy tensor, 
\begin{align}
  \label{eq:19}
  \mathcal{L}=-\frac{1}{M_{Pl}} h_{\mu\nu}^{(0)}T_{\mu\nu}  -\frac{1}{\Lambda_n} h_{\mu\nu}^{(n)}T_{\mu\nu}.
\end{align}
The coupling of each mode $n$ is given by the value of the graviton
wavefunction at the TeV brane; the wave functions are localized close to the $\tev$ brane so the
couplings are much stronger than $M_{Pl}^{-1}$,
\begin{align}
  \label{eq:6}
  \frac{1}{\Lambda_n}  =
\frac{|\alpha| ^{1/2}}{M_5^{3/2}}\frac{1}{|\alpha
    r_c|^{1/2}}\l(\frac{4n^2\pi^2}{4n^2\pi^2 +(\alpha
    r_c )^2}\r)^{1/2}.
\end{align}
For large mode number $n$, the spacing of the modes is $\delta m=
|\alpha|\pi/|\alpha r_c|\sim |\alpha|/ 20$. The coupling is mode-dependent
and is additionally suppressed from $(\tev)^{-1}$ by $(|\alpha|/M_5)^{1/2}$:
\begin{align}
  \label{eq:4}
  \Lambda_1 \, \, \quad &=\, \, 
  (80\tev){\l(\frac{|\alpha|}{M_5}\r)^{1/2}}{\l(\frac{M_{5}}{\tev}\r)^{-1}}  \\
\Lambda_{n>100} &= \, \,  (8
\tev)\,{\l(\frac{|\alpha|}{M_5}\r)^{1/2}}{\l(\frac{M_{5}}{\tev}\r)^{-1}}.
\end{align}

\subsection{Detection}

For large enough values of $\alpha$, the spacing between modes
($\delta m \sim |\alpha|/20$) becomes larger than the invariant mass
resolution of the LHC detectors. In this region of parameter space,
experimental searches for KK resonances have potential reach.  This is
a true jackpot scenario --- the modes are closely spaced, so upon
seeing one resonance, the LHC would be able to detect a multitude of
modes. The first hundred have a varying coupling strength and mass
spacing, and once distinct resonances become visible, all the
kinematically accessible modes will be seen above background after a
factor of $3$-$5$ increase in luminosity. This information, combined
with a measurement of the unique mass gap, can help pinpoint the extra
dimension as resulting from a Linear Dilaton background.

In particular, for $|\alpha|\sim 1\tev$, KK modes will have spacing
above the muon resolution of the ATLAS and CMS detectors and can be
seen as resonant peaks. At both detectors, muon tracks are
reconstructed with 97\% efficiency and an invariant mass resolution of
3-5\% between 400 GeV and 2 TeV, with 4\% at 1 TeV \cite{2010AIPC,
  Aad:2009wy}. An example invariant mass spectrum at the LHC in the dimuon channel is
shown in Figure~\ref{fig:invmass1000} for $|\alpha|=1\tev ,
M_5=3\tev$: already at a $7\tev$ LHC energy and $30\ifb$ of data,
individual resonances begin to become visible above the background and
detector resolution. For comparison, the low $\alpha$ spectrum in
Figure~\ref{fig:invmass100} would appear as a continuous excess at the
LHC detectors; it has the same energy dependence for any small value
of the curvature (including zero curvature).

\begin{figure}[t]
\centering
\subfigure[$|\alpha|=1\tev$ ]{
\includegraphics[trim = 0mm 81mm 0mm 88mm, clip, width=84mm]{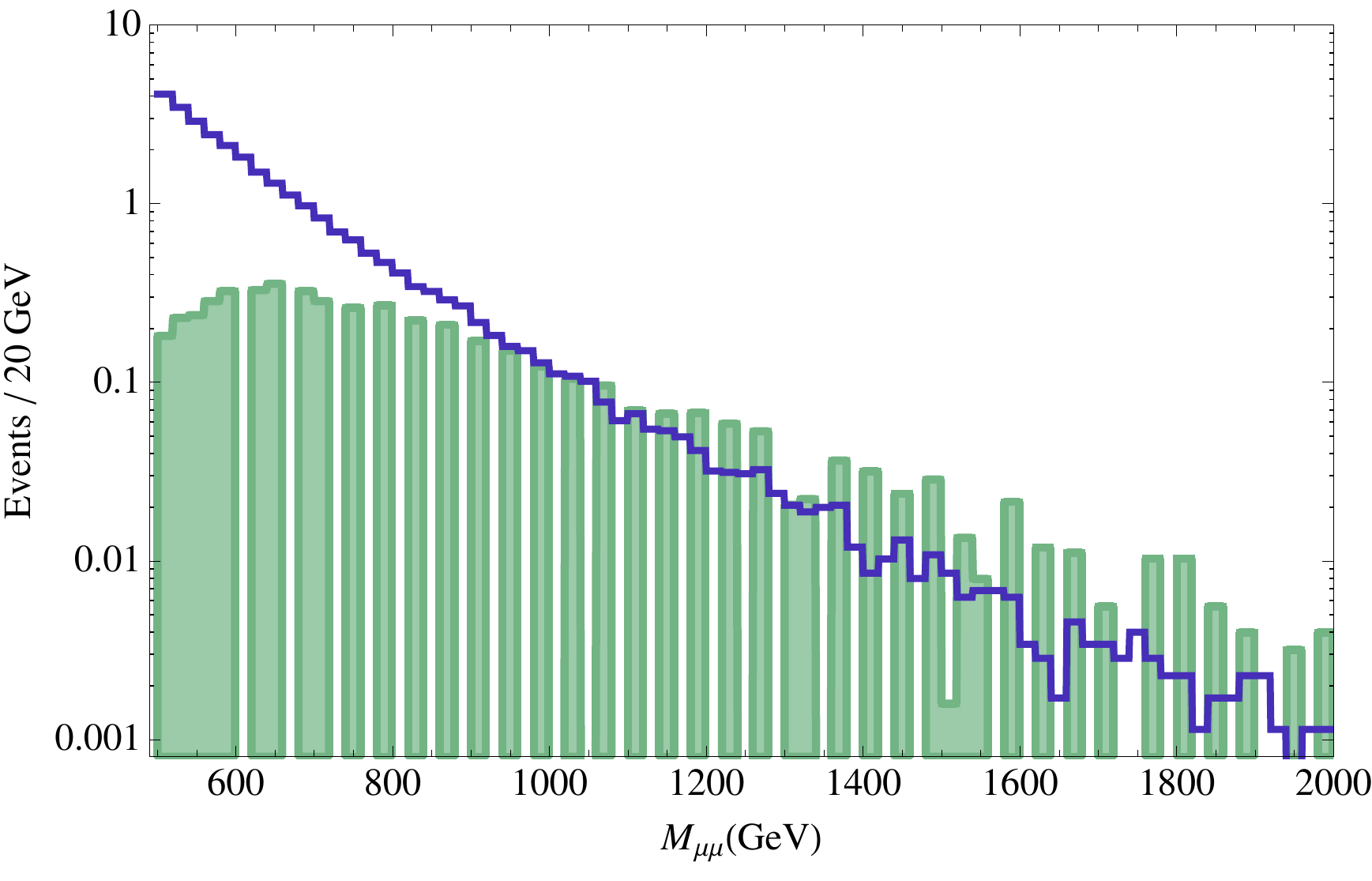}
\label{fig:invmass1000}
}
\hspace{-4mm}
\subfigure[$|\alpha|=100\gev$]{
\includegraphics[trim = 0mm 81mm 0mm 88mm, clip, width=84mm]{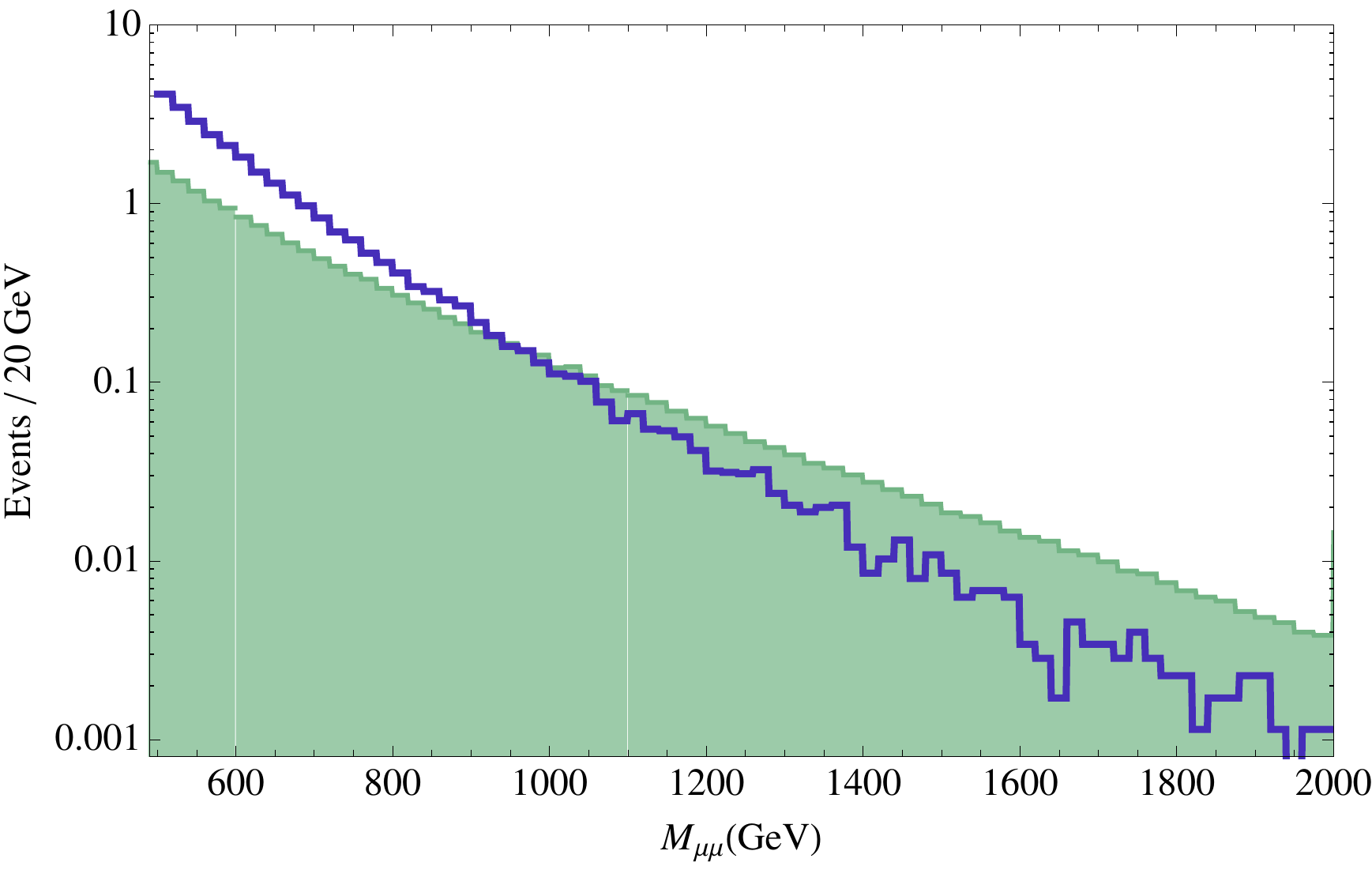}
\label{fig:invmass100}
}
\label{fig:mass_spectrum}
\caption[Optional caption for list of figures]{\small{The invariant mass
  spectrum at the LHC in the dimuon channel after $30\ifb$ integrated
  luminosity at $\sqrt{s}=7 \tev$ for example spectra with
  \subref{fig:invmass1000}\,$M_5=3\tev,|\alpha|=1\tev$, and
  \subref{fig:invmass100}\, $M_5=3\tev,|\alpha|=100\gev$. Green (gray):
  signal, blue (black): background. The data is grouped in $20\gev$ bins;
  detector resolution ranges from $12\gev$ at $400\gev$ to $40\gev$ at
  $1\tev$ and $100\gev$ at $2\tev$.  }}
\end{figure}

We can set bounds on this scenario from current LHC Randall-Sundrum
resonance searches, which depend on the coupling strength and mass of
each KK mode, as shown in Figure~\ref{fig:high_alpha}
\cite{Collaboration2011d,Aad2011d,Collaboration2011}.  These searches
are sensitive to KK gravitons decaying to pairs of photons or leptons
that can be seen as  invariant mass peaks above the background.  Similarly,
extrapolating from RS reaches gives the reach regions for high
$\alpha$ a center of energy $14\tev$ and integrated luminosity of
$100 \ifb$ shown in Figure~\ref{fig:high_alpha}.

\begin{figure}[t!]
\begin{center}
\includegraphics[trim = 0mm 81mm 0mm 80mm, clip, width=140mm]{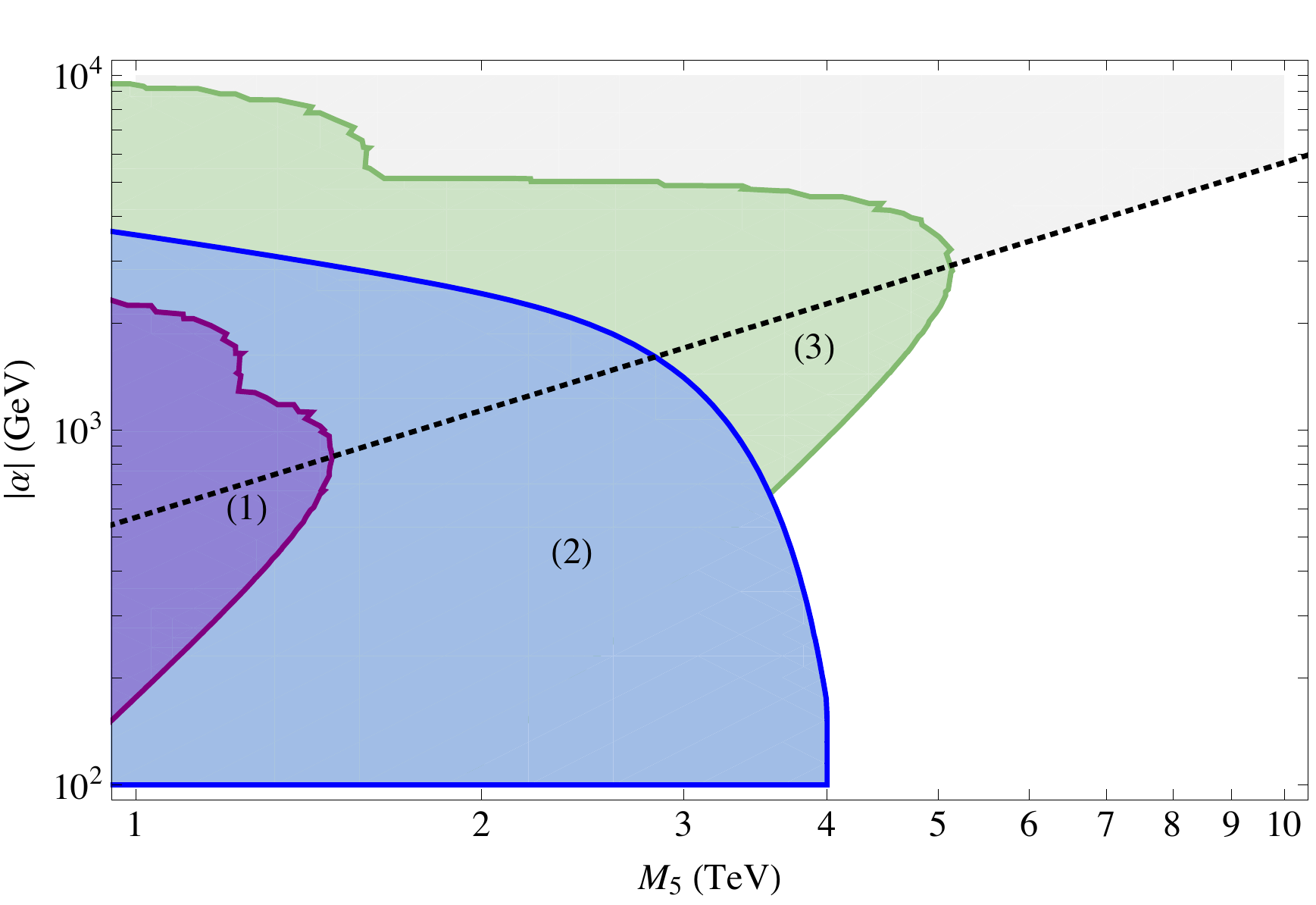}
\caption{\small{LHC limits on high curvature Linear Dilaton
  geometry. (1) Purple: current exclusion from RS resonance searches;
  (2) blue: current exclusion from ADD total cross-section searches;
  (3) green: LHC reach in resonance searches with $\sqrt{s}=14\tev$
  and $100$ fb$^{-1}$ integrated luminosity; gray (above dotted
  line): non-perturbative region where the curvature $|\alpha|$ becomes
  comparable to the 5-d Planck scale $M_5$. Reach in $M_5$ is limited
  by weak coupling, while reach in $|\alpha|$ is limited by production
  of the first mode of mass $|\alpha|/2$.}}
\label{fig:high_alpha}
\end{center}
\end{figure}

In addition to resonance searches, we consider bounds from the total
cross-section increase due to the extra modes. The dominant
contribution is from on-shell $s$-channel KK gravitons; the width of
each resonance is much smaller than the mass difference between
consecutive modes, and there is minimal interference with Standard
Model particles. We model the production and decay of KK modes up to
the cutoff $M_5$ for a range of $|\alpha|$ and $M_5$ in Madgraph 5.0
\cite{Alwall:2011uj} and compare the signal cross-section as a
function of center of mass energy with published limits
\cite{Aad2011d,Collaboration2011d,Collaboration2011,Dimensions2011a}. The
dependence on the cutoff is insignificant due to the extremely low
production rates at multi-TeV energies.

We find that we obtain more stringent bounds by including the highest
mass `control' region in addition to the signal region. We obtain a
limit $M_5 \gtrsim 3\tev$ for varying values of high curvature and no
limit for $|\alpha| > 4\tev$; the leading order $95\%$ CL limit is shown
in Figure~\ref{fig:high_alpha}.  The published information about
systematic uncertainties is only available for the highest mass signal
region so in calculating limits we have taken into account statistical
uncertainty only; thus our limit is an overestimate. A full
experimental analysis is required to obtain precise bounds.

\section{Low Curvature ($|\alpha| \,\ll \, M_5$ )}\label{LowAlpha}

\subsection{Physics of Low Curvature Linear Dilaton}
An interesting aspect of the linear dilaton geometry is that it
provides a natural framework for one large extra dimension with a
small curvature. In the limit where the linear dilaton slope goes to
zero, $\alpha \rightarrow 0$, the model reproduces ADD with one extra
dimension. From the 4-d point of view, a small
curvature $\alpha$ in the extra dimension lifts the spectrum of
Kaluza-Klein (KK) gravitons and avoids standard laboratory and
astrophysical constraints on large extra dimensions. In this section
we show that there are no bounds on a $\tev$ Planck scale from low
energy tests for $|\alpha|$ as low as tens of $\mev$.

On the other hand, the high energy LHC is sensitive not to low energy
details of the spectrum but to the entire discretum, so even with the
mass gap the low $\alpha$ spectrum will be resolved as one extra
dimension. Thus, there is potential to discover one large extra
dimension at the LHC, a possibility so neglected that it does not
appear in any experimental analyses. A similar phenomenological
scenario has been pointed out and analyzed previously for a low
curvature limit of RS in \cite{Giudice2005,Franceschini2011}. Still,
the phenomenology is distinct: there is a lower density of modes then
in ADD and each mode is more strongly coupled.  For instance, there
are no missing energy signatures because most KK modes have short
lifetimes and decay inside the detector.

In the following sections we present the leading bounds on low
$\alpha$ geometries, including emission in supernovae and BBN, fixed
target experiments, and the latest experimental bounds from the
LHC. The limits are shown in Figure~\ref{fig:low_alpha}.

\begin{figure}[t!]
\begin{center}
\includegraphics[trim = 0mm 80mm 0mm 80mm, clip, width=150mm]{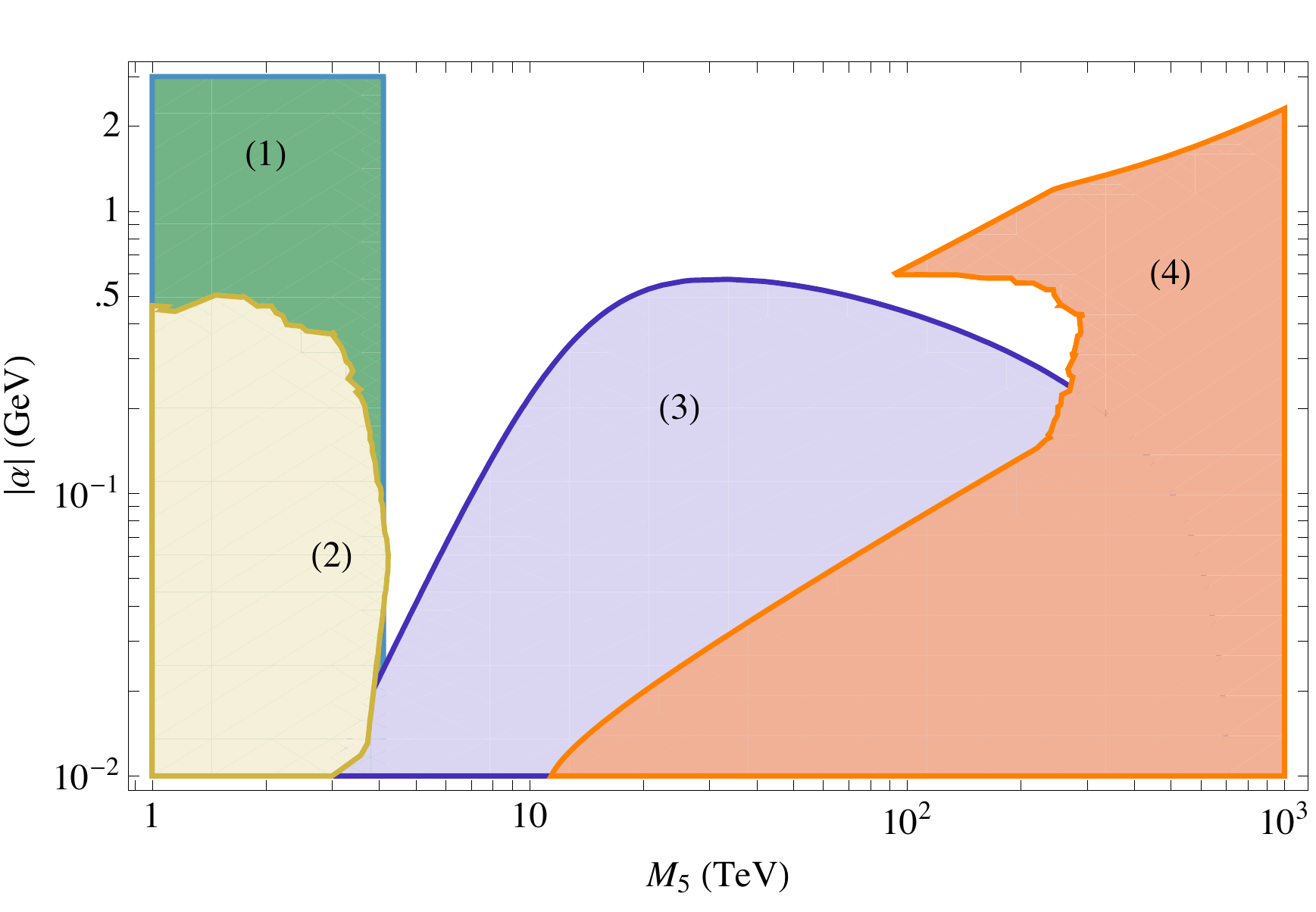}
\caption{\small{Current Bounds on Low Curvature Linear Dilaton. (1) Green:
  LHC limits at $\sqrt{s}~=~7\tev$ after $2.2\ifb$ integrated
  luminosity, (2) yellow: 10 events in the E137 electron beam dump
  experiment (zero events were observed over the course of the
  experiment), (3) purple: energy flux from Supernova 1987A, (4)
  orange: BBN electromagnetic and hadronic bounds. The upper lobe in
  BBN exclusion results from hadronic limits, with a lower threshold at
  $|\alpha|\sim 600\mev$ when hadronic decay channels become available
  to the lightest KK mode.}}
\label{fig:low_alpha}
\end{center}
\end{figure}

In the $|\alpha|\ll M_5$ limit we consider in this section, the
coupling $\Lambda_n^{-1}$ of each individual mode is extremely weak:
\begin{align}
  \label{eq:4}
  \Lambda_1 &= (8 \cross10^6\gev){\l(\frac{|\alpha|}{100\mev}\r)^{1/2}}{\l(\frac{M_{5}}{\tev}\r)^{-3/2}}\\
\Lambda_{n>100} &=  (8\cross10^5\gev)\l(\frac{|\alpha|}{100\mev}\r)^{1/2}\l(\frac{M_{5}}{\tev}\r)^{-3/2}
\end{align}
while the spacing of the modes is very small, 
\begin{align}
  \label{eq:46}
\delta m_n  &= 5\mev \l(\frac{|\alpha|}{100\mev}\r) \quad\quad\mathrm{for\,
  \,} n \gg 1.
\end{align}

Thus for many processes considered here, the contribution of a single
mode is undetectable and we will be interested in the total
cross-section where the entire discretum of modes participates in the process. To do
so we can approximate the sum over modes as an integral,
\begin{align}
  \label{eq:18}
   \sum_n = \int dm \,\frac{r_c}{\pi}\,\frac{m}{(m^2-\alpha^2/4)^{1/2}}.
\end{align}
Unlike the ADD and RS theories, in the linear dilaton geometry the
coupling of each mode is a function of mode number \eqref{eq:6}. To compute total
cross-sections we calculate the continuum integral of the
sum $\sum_n\sigma_n$. Since the coupling is very weak, the
leading production matrix elements contain single graviton couplings
and are proportional to $\Lambda_n^{-2}$,which gives
\begin{align}
  \label{eq:20a}
  \sum_n\Lambda_n^{-2}= \int dm\,  \frac{1}{\Lambda(m)^2} 
&=\,\frac{1}{\pi M_5^3} \int \,\frac{dm}{m}\,\l(m^2- \frac{\alpha^2}{4}\r)^{\half}.
\end{align}

In the $\alpha\rightarrow 0$ limit, the expression \eqref{eq:20a}
reduces to the sum over modes in one flat extra dimension, up to
a factor of $ 2\pi$ due to a difference in the definition of
$M_5$.\footnote[1]{The difference in conventions compared to flat extra
  dimensions appears in Eqn.~\eqref{eq:7}, where the length of the
  extra dimension is defined to be $r_c$ whereas it is often written
  as $2\pi R$.}. While the cross-section of a single
mode is proportional to ${|\alpha|}/{ M_5^{-3}}$, the production
cross-section for any mode is proportional to $M_5^{-3}$ and independent
of $\alpha$ given that all modes are accessible.

Another important quantity for understanding experimental signatures is the lifetime of the modes. The decay
rate is 
\begin{align}
  \label{eq:28}
  \Gamma_n &= \frac{m_n^3}{320\pi\Lambda_n^2}k(m),
\end{align}
where $k(m)$ is a function of the spin and number of open decay channels
($1$ for fermions, $2$ for bosons, etc.) \cite{Han1999}.
Even for a fixed number of decay channels, the lifetime varies sharply
with $\alpha$ and $M_5$, as well as mode number:
\begin{align}
\label{eq:32}
  &\tau_1 \,\,= \,(3\times10^{-5}\,\mathrm{s})
  \,\l(\frac{M_{5}}{\tev}\r)^3\l(\frac{|\alpha|}{100\mev}\r) ^{-4}\\
  &\tau_{100} = (4\times10^{-13}\,\mathrm{s})
  \l(\frac{M_{5}}{\tev}\r)^3\l(\frac{|\alpha|}{100\mev}\r) ^{-4}.
\end{align}
Lifetimes shorter than $10^{-4}$ s may be possible to measure in
laboratory experiments, while lifetimes longer than $1$ s can have an
impact on cosmology.

  We consider the parameter space of the theory in
the $(\alpha,\, M_5)$ plane (Figure~\ref{fig:low_alpha}). The 5-d
Planck scale $M_5$ determines the strength of the coupling of the
modes, while the curvature $\alpha$ sets both the mass spectrum and the coupling.

In sections \ref{astro} and \ref{beamdumps}, we present relevant
astrophysical and laboratory constraints; we find the first KK mode
can be as light as $10\mev$ in LD and $1$ MeV in low curvature RS and
avoid all existing bounds for TeV-scale $M_5$. Rare meson decays, such
as $K^+\rightarrow \pi^+ G$, could also potentially set limits on
$M_5$; however, as these processes are loop suppressed, the resulting
limits from meson experiments are much weaker than those due to fixed target
experiments \cite{Bijnens:2000wj}. In Section \ref{lhc} we compute the
LHC phenomenology and obtain a leading order limit of $M_5 \gtrsim 4.1
\tev$ and comment on optimal LHC searches and future reach of the
collider.

\subsection{Astrophysical Bounds}\label{astro}

\subsubsection{Supernovae}\label{Supernova}

One of the strongest bounds on the Planck scale in ADD comes from
overproduction of KK modes in Supernova 1987A
\cite{Hannestad2003,Hanhart2001,Cullen1999}. The
luminosity and duration of the SN1987A burst as measured by
the observed neutrino flux limits the energy loss due to other weakly
interacting light states to be less than $\dot\epsilon=10^{19}$ erg
g$^{-1}$ s$^{-1}$ at a typical supernova temperature of
$T=30\mev$ \cite{Raffelt:1999tx}. In the case of one flat extra
dimension, the density of light KK states is very high and results in
a strong limit: $M_5>7.4\cross10^5\gev$ \cite{Hannestad2003}. In the linear
dilaton geometry, a quick estimate gives the same limit for $|\alpha|
\ll T$ where the curvature is irrelevant, and no limit for
$|\alpha|/2>30\mev$, where the mass of first mode is above the SN
temperature. However, we find the thermally suppressed emission
of heavier gravitons as well as reabsorption in the SN core lead to a
more interesting constraint, as we discuss below.

\paragraph*{Graviton Emission}

The leading process for KK  production is nuclear gravistrahlung -- the
radiation of a graviton $G$ in the process of collision of two
nucleons $N$,
\begin{align}
  \label{eq:9}
  N+N\rightarrow N+N+G
\end{align}
Other processes such as $N+\gamma\rightarrow N+G$ and
$e+\gamma\rightarrow e+G$ are suppressed by $\alpha_{EM}$ and give
sub-leading contributions.  For soft graviton emission, the process
factorizes into the cross-section for nuclear scattering $\sigma_N$
and the probability for graviton emission $T^2\Lambda_n^{-2}$ where
$\Lambda_n^{-1}$ is coupling of the $n^{\text{th}}$ mode. For a non-degenerate,
non-relativistic nuclear medium of the neutron star, the energy loss
rate for a single graviton is given by \cite{Hannestad2003}
\begin{align}
  \label{eq:13}
  Q_m\propto \frac{1}{\Lambda(m)^2} \sigma_N n_B^2 T^{7/2}m_N^{-1/2}
\end{align}
where $\Lambda(m)$ is the graviton coupling, $T$ is the neutron star
core temperature, $n_B$ is the baryon density in the star, and $m_N$
is the nucleon mass. For typical neutron star conditions of $T=30\mev$
and $\rho = 3\cross 10^{14}$ g cm$^{-3}$, $\sigma_N = 25$ mb is a good
approximation for the nuclear cross-section.

We compute the total energy loss from a supernova through graviton
emission in the soft graviton approximation. For one large extra
dimension with radius $R$, the sum of the loss rates \eqref{eq:13}
over all modes approximated as an integral gives the energy loss rate $Q$ \cite{Hannestad2003}:
\begin{align}
  \label{eq:14}
  Q= \frac{2R}{(2\pi)^2}\int_0^{\infty}\,d\omega \,\omega\,S(-\omega)\,\int_0^{\omega}\,dm\,\l (\omega^2-{m^2}\r)^{1/2}\l(\frac{19}{18} +\frac{11}{9}\frac{m^2}{\omega^2} + \frac{2}{9}\frac{m^4}{\omega^4}\r)
\end{align}
where the integrals over mass of the mode $m$  and energy $\omega$
include the sum over graviton polarizations as well as integrals over
momentum and energy with dispersion relation $k=(\omega^2 -
m^2)^{1/2}$. $S(\omega)$ is a structure function up to first order
in $\omega/T$ which is derived using the single graviton energy loss
rate and the principle of detailed balance:
\begin{align}
\label{eq:22}
  S(-\omega) =
  \frac{1}{\omega^2}\frac{2}{1+e^{\omega/T}}\frac{1024\sqrt{\pi}}{5}\frac{\sigma_N
  n_B^2\,T^{5/2}}{m_N^{1/2}}\frac{1}{\Lambda_n^2}
\end{align}
Applying this analysis to the linear dilaton geometry, the larger mass
dependent coupling $\Lambda(m)^{-1}$ should be included in the integral
over modes. The mode spacing is also larger at $\pi/r_c \sim
\alpha/20$, but the spectrum can be still approximated as a continuum
starting at the lightest mode of $\alpha/2$ for $\alpha$ up to $1\gev$.  Then the expression for linear dilaton geometry is
\begin{align}
  \label{eq:24}
  Q &=\frac{S_0}{(2\pi)^2\pi M_5^3}\int_{\alpha/2}^{\infty}\,d\omega
  \,\frac{2}{1+e^{\omega/T}}\int_{\alpha/2}^{\omega}\,dm\,
  G({m}/{\omega},\alpha/2m)
\end{align}
where 
\begin{align}
  \label{eq:25}
  S_0=\frac{1024\sqrt{\pi}}{5}\frac{\sigma_N n_B^2\,T^{5/2}}{m_N^{1/2}}
\end{align}
and the integrand 
\begin{align}
  \label{eq:15}
  G(m/\omega,\alpha/2m) ={\l(1-\frac{m^2}{\omega^2}\r)^{1/2}}{\l(1-\frac{\alpha^2}{4m^2}\r)^{1/2}}\l(\frac{19}{18} +\frac{11}{9}\frac{m^2}{\omega^2}+ \frac{2}{9}\frac{m^4}{\omega^4}\r)
\end{align}
includes the mass-dependent coupling and reduces to the flat geometry
\eqref{eq:14} for $\alpha\rightarrow 0$. Evaluating the integral over
$m$ \eqref{eq:24} with typical SN conditions, $T=30\mev$ and $\rho =
3\cross 10^{14}$ g cm$^{-3}$, $\sigma_N = 25$ mb, and $n_B = 10^{-3}
\gev^3$ gives
\begin{align}
  \label{eq:26}
  Q    &=3.78\cross10^{41}
  \,\mathrm{erg\,\,cm}^{-3}\,\mathrm{s}^{-1}\,\l(\frac {T}{30\mev}\r)\l(\frac {M_{5}}{\tev}\r)^{-3}\mathcal{I}(\alpha/2T)
\end{align}
where the integral $\mathcal{I}(\alpha/2T)$ (with $x=\omega/T$),
\begin{align}
  \label{eq:27}
  \mathcal{I}(\alpha/2T)=\int_{\alpha/2T}^{\infty}\,dx
  \,\frac{x}{1+e^{x}}\frac{12}{\pi^2}\l(\frac{(\alpha/2T) ^6}{50 \,x^6}+\frac{(\alpha/2T)^4}{5\, x^4}+\frac{3 (\alpha/2T) ^2}{10\, x^2}-\frac{38 (\alpha/2T)}{25 \,x}+1\r)
\end{align}
is equal to $1$ for zero curvature and accounts for thermal
suppression of production for $\alpha>0$. For $\alpha<2T$, all modes are produced in nearly
equal number below the temperature of the supernova so production
is suppressed proportionally: $\mathcal{I}(0.1) =
0.9$ and $\mathcal{I}(0.5) = 0.5$ for example. If the lightest
mode is heavier than the temperature, then production is
suppressed exponentially.

To derive a bound on the geometry in terms of the $\alpha -M_5$ parameter
space, we impose the condition that the energy carried away by gravitons is
less than the total loss rate \cite{Raffelt:1999tx},
\begin{align}
  \label{eq:17}
  Q(\alpha/2T,M_5) &< \dot{\epsilon} \rho = 3\cross 10^{33} \,\mathrm{erg\,\,cm}^{-3}\,\mathrm{s}^{-1}.
\end{align}
For $\alpha/2 \ll T$, the curvature is irrelevant and the bound on
$M_5$ is the same as for $1$ flat extra dimension, $M_5 \gtrsim
7.4\cross 10^5 \gev$. For $\alpha/2 \lesssim T$, the production is
suppressed by the fraction of modes accessible below $T$. If the first
mode is heavier than the supernova temperature, production is
thermally suppressed and the limit on $M_5$ becomes much weaker; for
$\alpha=10T$, the suppression is $1/300$. Still, at $M_5$ of a few
$\tev$, the interaction cross-section is so large that even very
thermally suppressed KK towers with $m_n\gtrsim 0.5\gev$ are
experimentally excluded.

\paragraph*{Graviton Absorption}

When a KK graviton is produced and then scatters in the NS core, it
thermalizes and does not carry away energy. This process opens a
window in parameter space at low $M_5$ where the KK interaction
cross-section is highest.  At the same time, at low $M_5$, the
coupling is large and KK modes are overproduced by a factor of
$(7.4\cross 10^5\gev/M_5)^3$ at low $\alpha$, so all but a small
fraction of KK gravitons have to scatter before leaving the neutron
star to satisfy the energy loss bounds. 

The mean free path of the $n^{\text{th}}$  mode is $L_n= (\sigma n_B)^{-1}$ where $\sigma$ is the
interaction cross-section and $n_B$ is the baryon number density, $n_B
= 10^{-3} \gev^3$. The inverse gravistrahlung process,
$N+N+G\rightarrow N+N$, gives
\begin{align}
  \label{eq:12}
 L_n= (\sigma n_B)^{-1} = (\Lambda_n /T)^2 (25\, \mathrm{mb})^{-1}
 10^{3} \gev^{-3} = 200\, \mathrm{m} \,\l(\frac{M_{5}}{\tev}\r)^3\l(\frac{\alpha}{100\mev}\r)^{-1}
\end{align}

The density of the neutron star is very high so there is
negligible phase space suppression for the $3\rightarrow 2$ process.

The cross-section for absorption is proportional to $n^2$ for the
early modes so the first mode has the longest interaction length. Then
a conservative bound on the fraction of modes that escape the
supernova is $e^{-L_1/R_{SN}}$ where $R_{SN} \sim 10$ km is the radius
of the star. In the excluded region in Figure~\ref{fig:low_alpha}, we
take into account the varying interaction lengths weighted by the
thermal distributions of modes.  For the region in parameter space
relevant to reabsorption, KK decay length is longer than interaction length
and is not relevant.  The absorption rate is within a factor of a few
of more detailed estimates\cite{Hannestad2003}, which is the expected
accuracy of both calculations.

Another bound on KK gravitons from supernovae arises when gravitons
produced in a supernova are trapped in gravitational orbit around
the neutron star. As the gravitons gradually decay, the decay
products can hit the surface of the neutron star and reheat it, or in
the case of photons, be detected as gamma ray radiation
\cite{Hannestad2003}. However, for the parameter range considered here,
these bounds are irrelevant because all modes have lifetimes under one
year while the time scale of neutron star cooling is $10^5$ years or
more.  While very young neutron stars may have a residual density of
orbiting KK modes, these early environments are too noisy to allow detection
of trace amounts of additional radiation.

\subsubsection{Big Bang Nucleosynthesis}

Another thermal source of light KK gravitons is the early
universe. Gravitons can be produced in abundance at early times and
then decay to all kinematically allowed Standard Model particles,
affecting the delicate balance of Big Bang Nucleosynthesis (BBN).  The
standard model of nucleosynthesis matches measured relative
abundances of the light elements very well and is sensitive to 
processes that take place in the early universe between $1$-$10^6$
s. KK gravitons which decay too slowly and are too abundant can upset
this balance; we derive limits on the parameter space by requiring any
decay product interactions to be subdominant to established
rates at the time of decay.

The leading processes that affect BBN light element balance are proton
and neutron interconversion and deuterium over- or under-production relative to
He$^4$.  Electromagnetic $(e,\gamma)$ as well as hadronic
$(\pi,K,p,n,\ldots)$ decay products can contribute to these
interconversions. These processes have been used to constrain other
extra dimensional models as well as some supersymmetry models
\cite{Cyburt2003,Kawasaki2005,Jedamzik2006,Kawasaki2008,Bailly2009,Arvanitaki2005}. The
bounds depend on the lifetime of the KK graviton, the type and total
energy $E_d$ of the decay products, and the abundance $Y_G=n_G/s$ of
the gravitons.

For a given KK spectrum, the lifetime decreases rapidly with mode
number $n$ as $\tau\sim n^{-2}$ for low mass modes and $\tau\sim n^{-3}$
for higher modes, while the mass spacing is small, $\delta m\sim 0.1
m_1$, so the dominant bounds come from the first KK mode.

For $\alpha\lesssim 1\gev$, the light modes are very weakly coupled
and freeze-out while relativistic. Thus the abundance is not thermally
suppressed \cite{Kolb:1988aj},
\begin{align}
  \label{eq:1}
  Y_{G} = 0.278 \frac{g}{g_{*S}} \sim \mathcal{O}(1)
\end{align}
for each KK mode, where $g =5$ is the number of internal degrees of freedom of the
graviton and $g_{*S}$ is the number of effectively massless degrees of
freedom at freeze-out. Since the abundances are so high, there are
significant limits even for short lifetimes.

There are bounds on electromagnetic decay products starting at
$t\gtrsim10^4$ s from the observed ratio of D to He$^4$ and weaker
bounds from the He$^4$ to proton ratio $Y_p$ as early as
$t\gtrsim10^2$ s \cite{Cyburt2003, Kawasaki2005}. Since KK abundances
are order $1$, even these weak bounds can be used to constrain the
parameter space. We use the bound
$m_{G}Y_{G}\,e^{-t_{EM}/\tau}<10^{-6}\gev$ at $t_{EM}\gtrsim 10^2$ s
\cite{Kawasaki2005}. This limit is shown in orange in
Fig.~\ref{fig:low_alpha}.

For heavier modes, hadronic decays are the limiting factor: for a
particle of mass $100 \gev$ and dominant decays to hadrons, the bound
on energy deposited is $ m_G\,Y_G \lesssim 10^{-9}\gev$ at $t_H\gtrsim
0.1$ s\cite{Kawasaki2005}.  For KK masses
below $1\gev$, no nucleons are produced in decays. This weakens the
bound, but not entirely: for mesons at temperatures above $1\mev$, the
interaction length is comparable with the decay length, and a
$1$-$10\%$ of mesons induce charge exchange before decaying
\cite{Jedamzik2006}.  For decay times shorter than $1$ s, mesons
contribute to charge exchange, increasing He$^4$ abundance, with an
efficiency $\mathcal{E}\sim10\% $ of baryons \cite{Reno:1987qw}.

We impose $\mathcal{E}m_GY_G e^{-t_{H}/\tau}\lesssim 10^{-9}\gev$
to satisfy the bounds above. Heavier modes, $m_n\gtrsim\gev$, can
decay to baryons which have stronger interactions with
nuclei. However, the lifetimes of these modes are shorter and so the
effects on BBN are smaller than from mesons at relevant times.

Additionally, very light modes with lifetimes longer than one second
can change the number of relativistic degrees of freedom in the early
universe which contribute to the entropy density. Massless particles
have a significant effect on the rate of expansion of the universe
during BBN, leading to different freeze-out times of nuclear reactions
and different relative abundances of light species. Each KK graviton
mode has 5 degrees of freedom and can have a large effect; thus we
impose a hard limit of $\alpha \gtrsim 2\mev$ to exclude all masses below
$1\mev$ which would be effectively massless at the time of BBN.

For large extra dimensions, other cosmological bounds arise  from
overabundance of CMB photons or the over-closure of the
universe\cite{Hall1999}. These limits assume much longer lifetimes and are irrelevant
here.

\subsection{Fixed Target Experiments}\label{beamdumps}

Beyond astrophysical processes, we set direct limits on KK gravitons
from ground-based laboratory experiments. In particular, fixed target
beam dump experiments can be used to search for or rule out the
presence of light gravitons. Two types of beam dump
experiments---proton beam on target and electron beam on target---have
been designed to search for axions as well as to measure neutrino
oscillations \cite{MINOSCollaboration2011,Bays2011,Bjorken1988}. More
recently, interest in fixed target setups has been renewed by searches
for dark gauge forces \cite{Bjorken2009, Batell:2009di}. KK gravitons
can be produced in the beam collision with the target and travel
through shielding; their decay products can then be measured in a detector
downstream.

Proton beams on target---used in many generations of neutrino
oscillation experiments--- create $\pi$s that decay to
neutrinos. Gravitons can be produced in the collisions and have some
probability to travel underground and decay inside the Cherenkov
detector designed to detect neutrino interactions. The beam energy is
high at tens to hundreds of GeV leading to higher production rates,
but the probability for a graviton to decay in the volume of the
detector is small.

Electron beam dump experiments on the other hand were designed
specifically to search for decaying weakly coupled particles. The
target is followed by shielding to stop other beam products, followed
by a decay region. A detector measures decays of the weakly
interacting particles into pairs of electrons, photons, and in some experiments
muons.

To find current limits from beam dump experiments, there are two
quantities to calculate: the number of gravitons produced over the
duration of the experiment, and the probability that decay products of
a single graviton will reach the detector and pass the experimental
quality cuts.

The current leading limit comes from the E137 experiment at SLAC which
dumped a $20\gev$ electron beam onto a thin aluminum target followed
by 200 m of shielding and a 200 m decay region. Limits from other
electron beam dumps such as SLAC E141 and Fermilab E774 as well as
proton beams MiniBoone, MINOS are weaker than the combined E137 bound
and LHC bound discussed in Section \ref{lhc}.

New electron beam fixed target experiments have been proposed and some
are in the process of being implemented. The heavy photon search, on
schedule to begin data taking this year, may have a reach which
extends beyond the current bounds.

\paragraph*{Production}

The leading process for graviton production is gravistrahlung off the
electron beam mediated by virtual photons in the field of target
nuclei (Figure~\ref{fig:feynman}). The kinematics of the process at
graviton energy small compared to beam energy $E_G\ll E$ are the same
for any massive boson bremsstrahlung, and the model-dependent
$2\rightarrow 2$ process $e^-\gamma \rightarrow e^- G$ factors in the
limit of low momentum transfer. We follow the calculation in
\cite{Bjorken2009} to compute the differential cross-section for
graviton emission.
\begin{figure}[t!]
\begin{center}
\includegraphics[width=85mm]{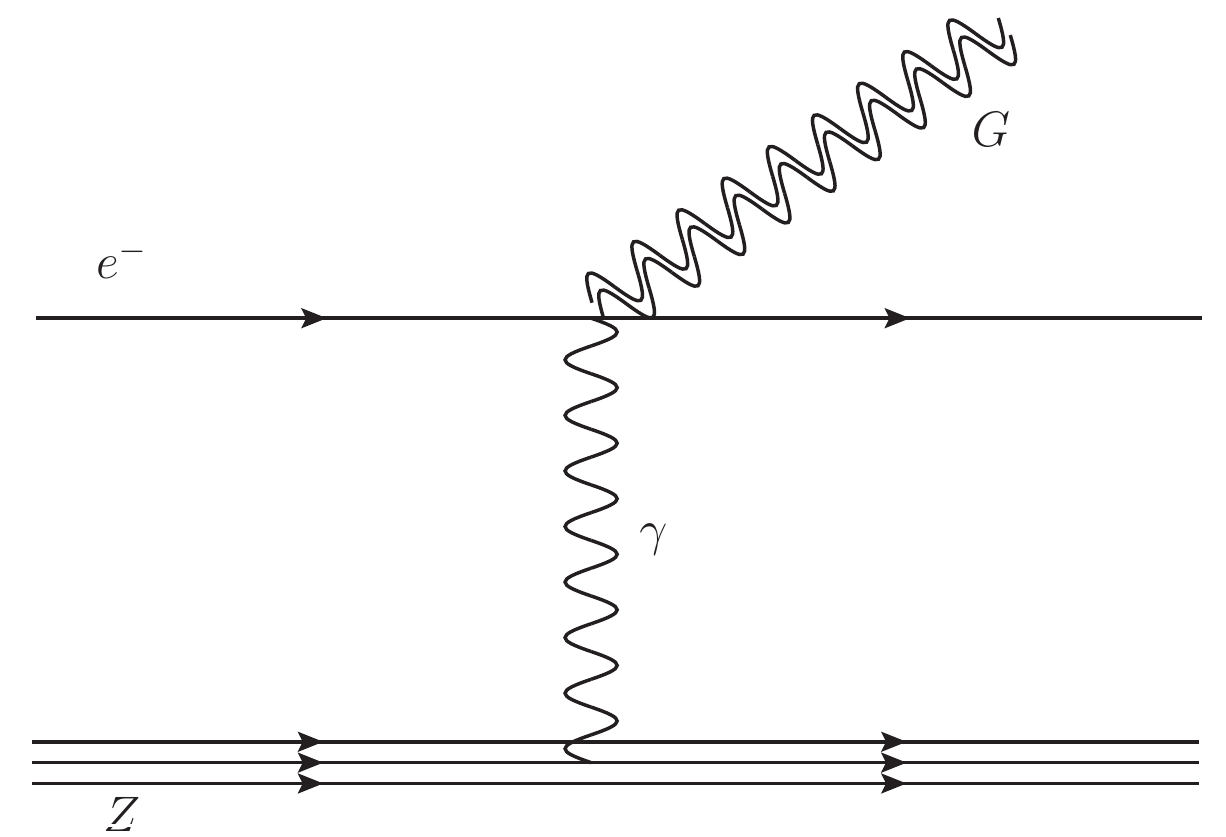}
\caption{\small{KK gravistrahlung off the electron beam in the field of target
  nuclei with atomic number $Z$.}}
\label{fig:feynman}
\end{center}
\end{figure}

The production cross-section is given by the Weizsacker-Williams
approximation \cite{Tsai:1986tx}, which relates the differential cross-section $\sigma$
for the $2\rightarrow 3$ process $e^-Z\rightarrow e^-\,Z\,G$
(Figure~\ref{fig:feynman}) to the $2\rightarrow 2$ process $e\gamma
\rightarrow eG$,
\begin{align}
  \label{eq:29}
  \frac{d\sigma}{dE_G\,d\cos\theta_G} &=\l. 2F_{\gamma}\frac{E \,x
  \beta_G}{1-x}\,\frac{d\sigma(e\gamma \rightarrow e G)}{dt}\r|_{min(-q^2)}
\end{align}
where $\beta_G =\sqrt{1-\frac{m_G^2}{E^2}}$, $x=E_G/E$ is the fraction
of the beam energy $E$ carried by the graviton, and $F_{\gamma} =
\frac{\alpha_{em}\,\chi}{\pi}$ is the effective photon flux sourced by
the target atoms and is a function of atomic number characterized by
$\chi\propto Z^2$. Here $q$ is the momentum of the photon and the
kinematics are evaluated at minimum momentum transfer in the
$2\rightarrow 3$ process.

The cross-section of the $2\rightarrow 2$ process for producing KK
mode n, $e^-\gamma \rightarrow e^-G$ is given by crossing symmetry
from the graviton production process $e^+e^- \rightarrow \gamma G$
\cite{Giudice1999},
\begin{align}
  \label{eq:2}
  \frac{d\sigma_{2\rightarrow 2}}{dt} &= \frac{\alpha_{em}}{16\Lambda_n^2\,s_{2\rightarrow 2}}F_2\l(\frac{t}{s},\frac{m_G^2}{s}\r)
\end{align}
where
\begin{align}
  \label{eq:5}
 F_1\l(x,\,y\r) &= \frac{-4 x (x+1) \left(2 x^2+2 x+1\right)+\left(16
      x^3+18 x^2+6 x+1\right) y+(4 x+1) y^3-6 x (2 x+1) y^2}{x
    (-x+y-1)} \nonumber\\
F_2\l(x,\,y\r) &= -(y - 1 - x)\,\, F_1\l(\frac{x}{y - 1 - x},\, \frac{y}{y - 1 - x}\r)
\end{align}

At minimum momentum transfer we can express \eqref{eq:29} with
$2\rightarrow 2$ cross-section \eqref{eq:2} in terms of $x,\,
m_e/m_G,\,\theta$, beam energy $E$ and graviton mass and coupling
$\Lambda_n^{-1}$. To evaluate the expression we expand in two small
parameters: the electron to graviton mass ratio $ m_e/m_G$ and the
angle of graviton emission relative to the beam $\theta \lesssim
(m_G/E)^{3/2}$. Integrating around $\theta =0$ and keeping only the
leading term in $\theta \rightarrow 0$ that gets regulated by either
the finite mass of the electron of the finite mass of the graviton,
$(m_G/E)^{3/2}$, we find the differential cross-section with respect
to graviton energy,
\begin{align}
  \label{eq:11}
    \frac{d\sigma_n}{dx} = F_{\gamma}\beta_G\frac{\alpha_{em}\left(1-2x+2 x^2\right) \log(1/\theta_{min}) }{8 \Lambda_n ^2}
\end{align}
Then integrating with respect to $x$ gives a total cross-section for
producing mode $n$
\begin{align}
  \label{eq:16}
  \sigma_n = \frac{\alpha_{em}^2\beta_G\chi}{12\pi\Lambda_n^2}\log(1/\theta_{min})
\end{align}

For the E137 experiment with a thin aluminum target and beam energy of
$20\gev$, $\log(1/\theta_{min})\sim 3$  and $\chi\sim 5 Z^2$, so
\begin{align}
  \label{eq:34}
    &\sigma_1 = 2\cross10^{-5} \,\mathrm{fb}\,\l(\frac{\alpha}{100\mev}\r)\l(\frac{M_{5}}{\tev}\r)^3\\
    &\sigma_{10} = 10^{-3}\,\mathrm{fb}\,\l(\frac{\alpha}{100\mev}\r)\l(\frac{M_{5}}{\tev}\r)^3
\end{align}

\paragraph*{Decay and detection}

Knowing the production cross-section and kinematics of the system, we
calculate the probability of decay products to enter the detector and
take into account detector acceptances.

The detector acceptance depends on the angular size of the
detector relative to the point $z$ where the decay occurs as well as
the energy fraction of the beam $xE$ of the KK graviton. For a thin
target like the one used in E137, the number of decays into with energy $xE$  at a point
$z$ after the target is given by \cite{Bjorken2009}
\begin{align}
  \label{eq:20}
  \frac{dN}{dxdz} = N_e \,n\, \frac{d\sigma_n}{dx}\l(e^{l_t/l_n} -
  1\r)e^{-z/l_n}\, \epsilon_{vis}
\end{align}
where $n $ is the number density of the target material, $l_t$ is the
target length, $l_n=\gamma v \tau_n$ is the decay length of the
$n^{\text{th}}$ KK mode in the lab frame, and $\epsilon_{vis}$ is the
graviton branching fraction into modes visible in the detector
(generally photons and electrons). For masses below $2m_{\mu}$,
$\epsilon_{vis}=2/3$ and for $m<300\mev$, $\epsilon_{vis}=6/11$ or
$8/11$ depending on whether the detector is equipped with muon
spectrometers.

The decay length depends strongly on the parameters and mode number:
 \begin{align}
  \label{eq:23}
 l_1&=  \l(4\cross
 10^6\mathrm{\,m}\r)\,\l(\frac{\alpha}{100\mev}\r)^{-5}\l(\frac{M_{5}}{\tev}\r)^3 \\
 l_{n>10}&= \l(4\cross 10^4\mathrm{\,m}\r)\l(\frac{10}{n}\r)^4\l(\frac{\alpha}{100\mev}\r)^{-5}\l(\frac{M_{5}}{\tev}\r)^3
\end{align}
The optimal decay length is on the order of the baseline of the
experiment; this allows for the largest production cross-section while
maximizing the chance for the decay products to enter the detector.

We sum over all modes $n$ to find the total number of expected
events. The excluded yellow region in Figure~\ref{fig:low_alpha}
corresponds to 10 events in the E137 detector; integrating
\eqref{eq:20} against detector geometry and acceptances produces a
$95\%$ limit given that no events were observed over the course of the
experiment.

\subsection{Large Hadron Collider}\label{lhc}
\subsubsection{Low Curvature Searches}\label{lhc_low_alpha}
Unlike systems sensitive to the low-mass and long-lived part of the KK
graviton spectrum such as supernovae, BBN, and beam dumps, the LHC
will be able to produce a multitude of states, possibly up to the
heaviest mode at the cutoff $\Lambda_{UV}$. For spacing between states
$\delta m$ smaller than the resolution of the detector, the Linear
Dilaton has many similarities at the LHC to one large flat extra
dimension. However, while current bounds on ADD can provide
approximate limits on the parameter space, they do not directly apply
since the experimental analysis for $1$ flat extra dimension has not
been performed.

There are several important differences between one large, low
curvature extra dimension and $\delta \geq 2 $ large extra
dimensions. First, the behavior of the cross-section as a function of
energy changes; the amplitude for $2\rightarrow 2$ $s$-channel
exchange processes with a KK intermediate state is
\begin{align}
  \label{eq:29}
  \mathcal{A} = \mathcal{S}(s)\l(T_{\mu\nu}T^{\mu\nu} - \frac{T_{\mu}^{\mu}T_{\nu}^{\nu}}{3}\r)
\end{align}
where
\begin{align}
  \label{eq:10}
  \mathcal{S}(s) = \frac{-i\pi}{M_5^3 \sqrt{s}} + \frac{2}{\Lambda_{UV}},
\end{align}
while for $n=2$ extra dimensions, $\mathcal{S}(s) \propto \ln
(s/\Lambda_{UV}^2)$, and for $n>2$, $\mathcal{S}(s) \propto
\Lambda_{UV}^{n-2}$ and is independent of $s$
\cite{Franceschini2011}. Based of the fact that in $n\geq 2$ extra
dimensions the signal cross-section does not fall with center of mass
energy, the experimental signal regions have been selected to be at
very high invariant masses
\cite{Aad2011d,Collaboration2011d,Collaboration2011,Dimensions2011a}.
For the linear dilaton geometry, the cross-section is inversely
proportional to $s$ and so the searches optimized for $\delta\geq 2$
are less effective. Furthermore, just for one extra dimension is the
cross-section only weakly sensitive to the cutoff; in the
remainder of the paper we take the cutoff to be $\Lambda_{UV} \simeq
M_5$.

Second, in low curvature models gravitons have a short lifetime and
can be produced on shell so the production cross-section is
enhanced. In the narrow-width approximation, the enhancement is by
a factor of $1/\epsilon$ where
\begin{align}
  \label{eq:43}
  \epsilon =\l.\frac{\pi\,\Gamma_n}{2\,\delta m_n}\r|_{m_n=\sqrt{s}}
  =\frac{k}{2\pi}\l(\frac{\sqrt{s}}{M_5}\r)^3 
\end{align}
and $k$ is an order one number that depends on the available decay
modes of the graviton into Standard Model particles
\cite{Franceschini2011}. In fact, since the width is still much
smaller than the mass splitting of states, on-shell production
dominates over interference with SM processes. This increases the
importance of processes with $s$-channel exchange, such as dilepton
and diphoton final states. 

Last, for the parameter range allowed by astrophysics and fixed
target experiments, a vast majority of states decay immediately inside
the detector, and at most a few tens of modes will have a displaced
vertex decay. Thus the jet plus missing energy search which currently
places the tightest bounds on one extra dimension, is no longer
relevant. There is a small signal from the decay of gravitons to
neutrinos, but the branching fraction is less than $4\%$. Since the
cross-section scales as $M_5^{-3}$, this weakens the bound on $M_5$
from $j+\slashed{E}_T$ searches by a factor of $1/3$.

A fraction of a percent of the modes for $\alpha$ on the order of
several GeV will decay inside the detector with displaced vertices
between $100~\mu$m and several meters. While the displaced vertex
signature is spectacular, it is a poor search channel: the total
cross-section for displaced vertices are $1$ fb or less while
displaced vertex searches are currently sensitive to cross-sections of
$1$-$100$~pb and are tailored to supersymmetry signals, enforcing
additional trigger requirements \cite{Jackson:2011zi}.

Without a missing energy or displaced vertex handle for searches, the
experiments are limited to searches for an overall excess above
Standard Model backgrounds in channels corresponding to graviton decay
products: dijets, dileptons (muons and electrons), and
diphotons. Current experimental searches focus on cut-and-count
measurements in the invariant mass spectra: with the assumption that
background falls more rapidly than signal with increasing center of
mass energy, the signal region is simply defined with a lower bound on
the dimuon (diphoton, dijet) invariant mass. Then limits are set by
comparing the observed and expected event counts in the signal
region. Current searches have signal regions start at an invariant
masses near $1\tev$, depending on the channel
\cite{Aad2011d,Collaboration2011d,Collaboration2011,Dimensions2011a}.

Gravitons couple universally to the stress-energy tensor so branching
fractions to SM particles only depend on the degrees of freedom of the
decay products. Due to color factors, the leading decay is to jets;
dijet final states are dominated by $t$-channel $uu\rightarrow uu$.
However, even though the $pp\rightarrow G \rightarrow jj$ rate is
higher than diphoton or dilepton, the sensitivity and reach to $M_5$
is worse because of large systematic uncertainties on the 
normalization and shape of jet distributions \cite{Giudice2005}.

Potential reach in the dijet channel can be improved by considering
observables with less systematic uncertainty, such as the angular
distance between two final jets as proposed in
\cite{Franceschini2011}. However this analysis is least effective for
one extra dimension, where the angular dependence is most similar to
the Standard Model, resulting in a weak bound: $M_5>0.8\tev$ for the
latest ATLAS results with $36$ pb$^{-1}$
\cite{Franceschini2011,Aad2011b} . If the experimental collaborations
release updated analyses, these limits may become competitive with
current bounds from dileptons and diphotons and can provide another
tool for characterizing the theory.
 
The cleaner dimuon and diphoton cut-and-count searches currently
provide the most stringent bound on the low scale of gravity, $M_5$.
Gravitons have comparable decay rates to muons and photons,
$\Gamma(G\rightarrow\gamma\gamma) =2\Gamma(G\rightarrow\mu\mu) $,
while Standard Model processes produce many more dimuons from
Drell-Yan than diphotons. Thus, the bound to photons is stronger
despite lower photon detection efficiencies and large uncertainties in
the diphoton mass spectrum. Electron reconstruction has similar
uncertainties to that of photons but electrons are produced in
abundance through DY and so provide a weaker limit than either
dimuons or diphotons.

To convert current experimental limits to bound on $M_5$ in the LD
scenario, we model a generic small-$|\alpha|$ model at the LHC and
derive a signal cross-section $\sigma_s$ at invariant masses up to
$M_5$. For a high-energy collider, the 5-d curvature is irrelevant as
long as $|\alpha|\ll M_5\sim \tev$.  This is due to the fact that high
momentum is insensitive to curved space for $p^2\gg \alpha^2$; thus
high-$n$ modes effectively see a flat extra dimension and have equal
couplings. In the high-energy limit, both LD and low curvature RS KK
spectra depend only on the 5-d Planck scale and the size of the
``box'' in which the gravitons propagate. The production
cross-sections at high $n$ are $\sigma_n\propto (r_c M_5^3)^{-1}$ and
the models have identical phenomenology. 

Using this intuition, we model a production scenario at the LHC with
low curvature $|\alpha|$ as a sum over modes with $|\alpha| = 100\gev$
up to a cutoff at $M_5$. The first mode, at $50\gev$, is below most
detector energy cuts and the spacing between modes ranges from $0.1$
to $5\gev$, well below the resolution of the detector.  The events are
generated in MadGraph 5.0 \cite{Alwall:2011uj}. We verify that our
results do not depend on the value of $\alpha$ at energies above
$|\alpha|/2$. The decay width $\Gamma \ll \delta m$, and so there is
no interference between the modes or with Standard Model particles in
the $s$-channel. The dimuon invariant mass spectrum at the LHC for a
background and signal example spectrum with
$M_5=3\tev,\,|\alpha|=100\gev$ after $30\ifb$ integrated luminosity at
$\sqrt{s}=7\tev$ is shown in Figure~\ref{fig:invmass100}.

We consider the published bounds on signal cross-sections from
experimental collaborations at $2.2\ifb$ integrated luminosity,
$\sigma_s<1.2$ fb for $m_{\mu\mu}>1100\gev$ and $\sigma_s<3.0$ fb for
$m_{\gamma\gamma}>900\gev$
\cite{Dimensions2011a,Collaboration2011}. We take into account
detector acceptances, cuts, and approximate NLO K-factor corrections,
and find a bound from $pp\rightarrow\mu\mu$ to be $M_5>2.7\tev$ and a
stronger bound $pp\rightarrow\gamma\gamma$ to be $M_5>3.1\tev$ for
$K=1.6$ and $M_5>2.6\tev$ for $K=1.0$. A summary of the limits is
shown in Table~\ref{M5limits}.

\begin{table}[t!]
\begin{center}
 \begin{tabular}{ | c | c || c | c | c| }
    \hline
& Diphotons & & Dimuons \\\hline
 K factor  & Signal Region & K factor  & Signal Region  \\  
                  & $M_{\gamma\gamma} > 900 $ \quad $M_{\gamma\gamma} > 500$ & &$M_{\mu\mu} > 1100 $ \quad $M_{\mu\mu} > 600$ \\ \hline \hline
    $ 1.6$ & $ 3.1 \tev $  \quad  $ 4.8 \tev$ &$ 1.3$ & $ 2.7 \tev $  \quad  $ 3.5 \tev$  \\ 
    $ 1.0$ & $2.5\tev$ \quad $4.1\tev$ & $ 1.0$ & $2.5 \tev$ \quad $3.2\tev$ \\
    \hline
  \end{tabular}
\end{center}
\caption{Current 95\% CL limits on $M_5$ at $7\tev$ after $2.2\ifb$ in
  diphoton and dimuon searches.}
\label{M5limits}
\end{table}

However, we find that as the searches were optimized for $\delta\geq
2$ large extra dimensions, the tighter bounds would actually come by
including some of the `control' regions with lower invariant
mass. From the published 7 TeV CMS studies
\cite{Dimensions2011a,Collaboration2011}, we calculate an approximate
exclusion from the $2.2\ifb$ diphoton spectrum of $M_5\geq 4.1\tev$
for leading order production (K factor of $1.0$) and $4.8\tev$ for NLO
production $(K=1.6)$ at $95\%$~CL. We find a sub-leading bound of
$M_5\geq 3.2\,(3.5)\tev$ for $K =1.0\,(1.3)$ at $95\%$~CL from the
dimuon spectrum. The leading order graviton production bound of
$4.1\tev$ is shown in Figure~\ref{fig:low_alpha}. Without detailed
knowledge of systematic uncertainties for these signal regions, our
limits are an overestimate. Accurate limits can only be set by the
experimental collaborations with a reconsideration of the signal and
control regions and the systematic uncertainties at different
energies.

\subsubsection{Future Reach}

The most promising avenue for exploring the low-curvature parameter
space is the Large Hadron Collider. In this section we perform an
analysis similar to that in Section~\ref{lhc_low_alpha} to find the
ultimate reach in $M_5$ for low $\alpha$ with a center of mass energy
$\sqrt{s}=14\tev$ LHC.  The best limits come from diphoton invariant mass
measurements; dimuon searches are less effective but can provide an
important confirmation in case of discovery.

We calculate signal and background spectra at $\sqrt{s}=14\tev$ with
MadGraph 5.0 \cite{Alwall:2011uj}.  For the diphoton channel, in
addition to diphoton production $pp\rightarrow \gamma\gamma$, we
approximate the contribution from jet conversion or misidentification
in the $\gamma$ + jet and jet-jet channels as an up to $10\%$ effect
for $m_{\gamma\gamma} >900\gev$. For concreteness we assume current
experimental parameters: $22\%$ background uncertainty, signal
efficiency of $76.4 \pm 9.6\%$, a background $K$ factor of $1.3$, and
the same isolation and $\eta$ cuts as the current 7 TeV CMS studies
\cite{Collaboration2011}.  Optimizing the reach for
$10 \ifb$ gives a signal region $m_{\gamma\gamma} >1000\gev$, and for
$100 \ifb$ a signal region $m_{\gamma\gamma} >1500\gev$.

For the dimuon channel we again assume current experimental
parameters: $16\%$ background uncertainty , signal efficiency of
$90\%$,and the same isolation and $\eta$ cuts as the current 7 TeV CMS
studies \cite{Dimensions2011a}.  In this case we
find an optimal signal region $m_{\mu\mu} >1100\gev$ at $10 \ifb$, and
$m_{\mu\mu} >1400\gev$ at $100 \ifb$.

For small curvature, although the jet plus missing energy channel
would give the best reach for one extra dimension (up to $17 \tev$ for
$100 \ifb$), without the missing energy signature the NLO reach is
$M_5\gtrsim 7.9 \tev$ for $10 \ifb$ and $M_5\gtrsim 9.0 \tev$ for $100
\ifb$. The detailed results for $\sqrt{s}=14\tev$ are listed in Table~\ref{M5reach}.

Without detailed experimental background studies at  $\sqrt{s}=14\tev$ the
limits are only approximate.  Nevertheless, the total signal
cross-section $\sigma_s$ scales as $M_5^{-3}$ so our limits are
only weakly sensitive to errors in background modeling or systematics: a change
in the background cross-section in the signal region by $50 \%$
results in a $10\% $ change in $M_5$ reach.

\begin{table}[t]
\begin{center}
 \begin{tabular}{ | c | c || c | c | c| }
    \hline
& Diphotons & & Dimuons \\\hline
 K factor  & Integrated Luminosity & K factor  & Integrated Luminosity  \\  
                  & $10 \ifb $ \quad $100 \ifb$ & &$10 \ifb $ \quad $100 \ifb$ \\ \hline \hline
    $ 1.6$ & $ 7.9 \tev $  \quad  $ 9.0 \tev$ &$ 1.3$ & $ 6.6 \tev $  \quad  $ 7.8 \tev$  \\ 
    $ 1.0$ & $6.7 \tev$ \quad $7.7\tev$ & $ 1.0$ & $6.1 \tev$ \quad $7.1\tev$ \\
    \hline
  \end{tabular}
\end{center}
\caption{Future reach: 95\% CL limits on $M_5$ for $|\alpha| \ll M_5$
  at the  $\sqrt{s}=14\tev$ LHC.}
\label{M5reach}
\end{table}

\section{Low Curvature Randall-Sundrum}\label{LowRS}

For low curvature Randall-Sundrum models, the same sets of limits
apply as in the low curvature linear dilaton model. We apply our
analysis of Sections~\ref{astro}--\ref{lhc} to the scenario presented
in \cite{Giudice2005} and consider the two-parameter space of the 5-d
Planck scale $M_5$ and first KK mass $m_1$
(Figure~\ref{fig:low_alphaRS}). In this case, the coupling is
independent of mode number and is given by
\begin{align}
  \label{eq:47}
\Lambda&= (2.8
 \cross10^5\gev){\l(\frac{m_1}{50\mev}\r)^{1/2}}{\l(\frac{M_{5}}{\tev}\r)^{-3/2}}
\end{align}
and the KK mass spectrum is $  m_n = \dfrac{x_n }{x_1}\,m_1$, where
\begin{align}
  \label{eq:8}
  x_n& = \beta_n -
\frac{3}{8\beta_n}+\frac{3}{128\beta_n^3}-\frac{1179}{5120\beta_n^5}+\ldots,\qquad
\beta_n = \l(n+\frac{1}{4}\r)\pi
\end{align}

Unlike the linear dilaton geometry, in low curvature RS the spacing
between KK modes is parametrically the same as the mass gap $m_1$; for
a fixed $m_1$ there are fewer modes up to the cutoff than in LD and
each mode is more strongly coupled than in LD. In the case of
BBN for example, the contributions of the first several modes to the
hadronic limit are visible individually (Fig.~\ref{fig:low_alphaRS}).
\begin{figure}[t!]
\begin{center}
\includegraphics[trim = 0mm 80mm 0mm 80mm, clip, width=150mm]{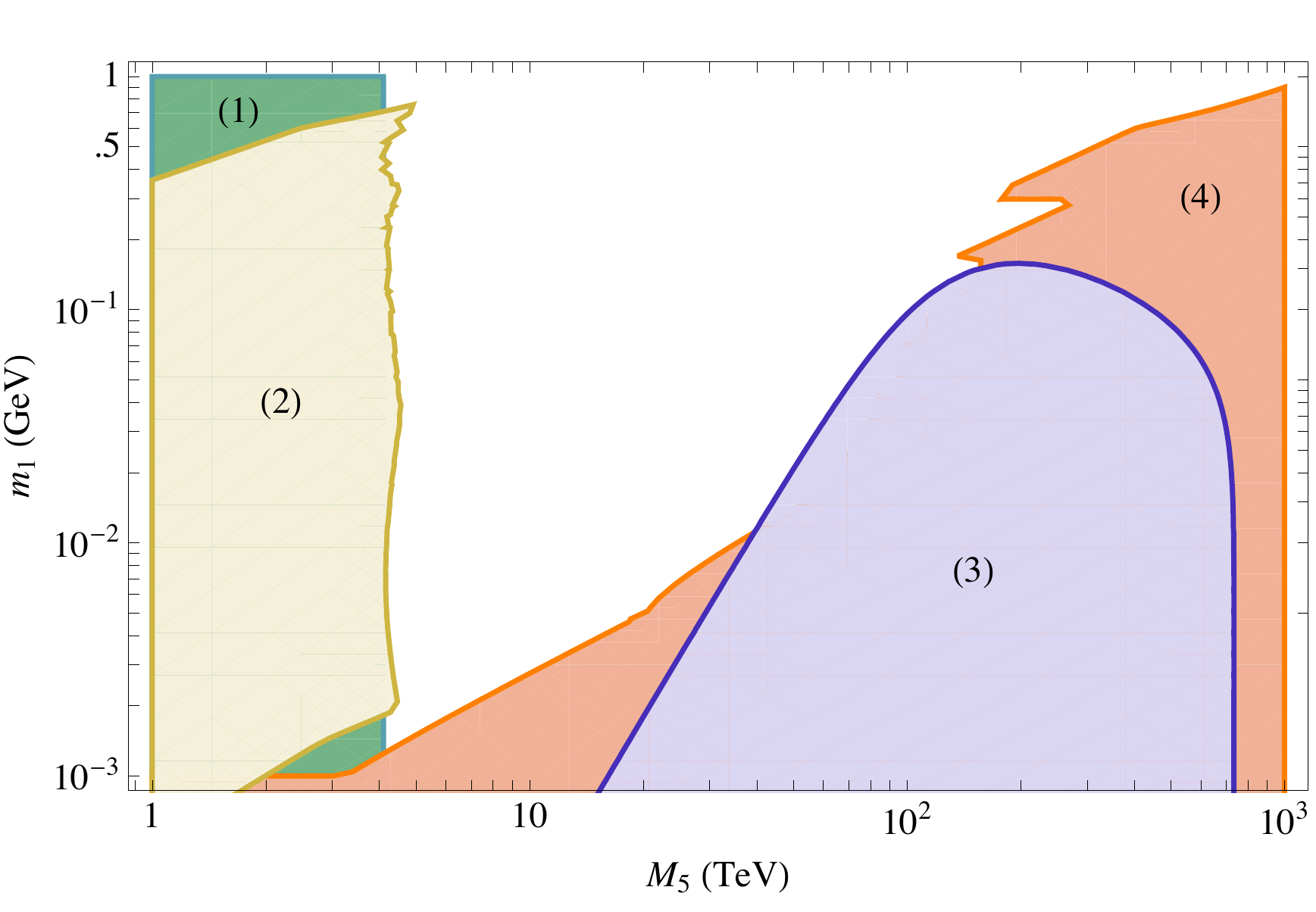}
\caption{\small{Current Bounds on Low Curvature RS.  (1) Green:
  LHC limits at $\sqrt{s} = 7\tev$ after $2\ifb$ integrated
  luminosity, (2) yellow: 10 events in the E137 electron beam dump
  experiment (zero events were observed over the course of the
  experiment), (3) purple: energy flux from Supernova 1987A, (4) orange: BBN electromagnetic,
  hadronic, and entropy bounds.}}
\label{fig:low_alphaRS}
\end{center}
\end{figure}

Since each KK mode in low curvature RS is more strongly coupled than
the corresponding mode in LD, the reach with beam dump experiments
extends to higher curvature. In contrast, the shorter decay lengths
and stronger couplings significantly weaken bounds from supernovae and
BBN: there are no constraints from astrophysics at TeV-scale $M_5$
down to a mass gap of $ 1\mev$. The LHC on the other hand is
insensitive to the details of the spectrum, so the collider bounds and
future reach (Table~\ref{M5reach}) on $M_5$ are identical in low
curvature RS and LD models.

\section{Discussion and Conclusions}\label{conclusions}

We have studied the phenomenology of an extra dimension with a linear
dilaton background, inspired by possibilities of Little String
Theory at a TeV. The linear dilaton background gives rise to a unique
gapped mass spectrum, where the curvature $|\alpha|$ in the extra
dimension lifts the graviton KK spectrum. The corresponding radion
phenomenology is studied in a complementary paper \cite{Tony}. Other
states may also result from LST at a TeV; these depend on the
compactification details and can be strongly coupled, so we do not
consider them here \cite{Antoniadis2011}.

For large curvatures $|\alpha|$ close to the five dimensional Planck
scale $M_5$, the LHC can provide a wealth of information. KK modes are
separated by more than the resolution of the LHC, and if one resonance
is detected it will be possible to study a tower of light resonances
by the end of the LHC run.  The mass gap of size $|\alpha|/2$ and
mode-number-dependent coupling $\Lambda_n$ can identify an excess at
the LHC as resulting from the linear dilaton scenario. The resonances
can be clearly identified as gravitons by studying branching fractions
to muons, photons, electrons, and jets, which are set by the universal
coupling of the KK modes to the stress-energy tensor. In addition the
spin of the resonances can be established with sufficient statistics
by measuring angular correlations of decay products \cite{Hewett2002}.

Another possibility is one large extra dimension at the LHC: for low
curvatures $|\alpha| \ll M_5$, the LHC is insensitive to details of
the graviton spectrum. In this region of parameter space the
contribution of KK modes will appear as a continuous excess with the
same dependence on center of mass energy as a large flat extra
dimension.  As we demonstrate in Section~\ref{LowAlpha}, the small
mass gap avoids exclusions from supernovae and BBN, and as each KK
mode is much more strongly coupled than $M_{Pl}^{-1}$, the gravitons
decay in the detector and do not carry away missing energy. This
possibility has not been included in experimental analyses; we
calculate the current limit based on published data and demonstrate
that as the searches have not been optimized with one large extra
dimension in mind, the reach of the searches could be improved by
relaxing the invariant mass cut of the signal region.

The phenomenology of low curvature RS models is similar to that of the
low curvature linear dilaton, and we apply our analysis to this case
as well. We find that the first mode of the theory can be as light at
$1\mev$ and avoid astrophysical limits. The best avenues for discovery
are new beam dump experiments and the LHC at increased center-of-mass
energy.

The most promising discovery channel at the LHC is the diphoton
invariant mass spectrum due to low Standard Model backgrounds at high
energies, followed by dimuons. For large curvatures, a $14 \tev$ LHC
can reach $M_5\sim 5\tev$ depending on the strength of the coupling
set by $|\alpha|$. For small curvature, the reach goes up to $7.9
\tev$ for $10 \ifb$ and $9.0 \tev$ for $100 \ifb$.

If a signal corresponding to one large extra dimension is observed at
the LHC, one exactly flat extra dimension can clearly be excluded by
lack of missing energy signatures. Further information about the
detailed geometry will require low-energy experiments, since without
sensitivity to individual resonances it is impossible to distinguish
low curvature theories. For a sufficiently low 5-d gravity scale,
fixed target experiments could detect individual, longer lifetime
states; for example, the currently commissioned Heavy Photon Search
proposed to search for light dark matter \cite{Bjorken2009, HPS} can
study coupling strengths and lifetimes of KK gravitons with masses of
up to several GeV.

\section{Acknowledgements }
We thank Ignatios Antoniadis, Savas Dimopoulos, Amit Giveon, Natalia
Toro, and especially Asimina Arvanitaki for useful discussions.  MB is
supported in part by the NSF Graduate Research Fellowship under Grant
No. DGE-1147470. This work was supported in part by ERC grant
BSMOXFORD no. 228169.

\bibliography{GravitonPhenomenologyofLinearDilatonGeometries}
\bibliographystyle{h-physrev}

\end{document}